\documentclass{aastex62}

\shorttitle{An Additional inner planet in Kepler1647?}
\shortauthors{Hong et al.}

\begin{document}

\title{\uppercase{Could there be an undetected inner planet near the stability limit in Kepler-1647?}}

\author{Ziqian Hong}
\affil{Center for Relativistic Astrophysics, School of Physics, 
Georgia Institute of Technology, Atlanta, GA 30332, USA}
\affil{Department of Astronomy, University of Science and Technology of China, Hefei, Anhui 230026, China}

\author{Billy Quarles}
\affil{Center for Relativistic Astrophysics, School of Physics, 
Georgia Institute of Technology, Atlanta, GA 30332, USA}

\author{Gongjie Li}
\affil{Center for Relativistic Astrophysics, School of Physics, 
Georgia Institute of Technology, Atlanta, GA 30332, USA}

\author{Jerome A. Orosz}
\affil{Department of Astronomy, San Diego State University, 
5500 Campanile Drive, San Diego CA 92182, USA}



\begin{abstract}

Kepler-1647b is the most recently discovered planet that transits two stars, i.e., a circumbinary planet (CBP). Due to its large orbital separation, Kepler-1647b stands out from the rest of the Kepler CBPs, which mostly reside on much tighter orbits near the stability limit. The large separation of Kepler-1647b challenges inward disk migration as a dominant formation pathway, suggested by the other Kepler CBPs. In this paper, we consider the possibility of an undetected planet near the stability limit by examining observational consequences of such a planet. We calculate the transit probability of the putative planet, transit timing variations (TTVs) of the known planet, and eclipsing timing variations (ETVs) of the host binary caused by the putative planet. We find the presence of a $\gtrsim$30M$_{\oplus}$ inner planet to be highly unlikely near the stability limit. In addition, we provide future TTV observation windows, which will further constrain possible undetected planets with lower masses.

\end{abstract}



\section{Introduction} \label{sec:intro} 

The discovery and characterization of circumbinary planets (CBPs) have greatly contributed to our understanding of planet formation in short period stellar binary systems. Our knowledge of CBPs is mainly derived from the photometric observations from the Kepler Mission which have provided unambiguous detections of CBPs and is in contrast to prior observations of post common envelope binary systems \citep[e.g.,][]{Beuermann2010}. So far, eleven Kepler transiting CBPs have been discovered, including Kepler-16b \citep{Doyle2011}; Kepler-34b and 35b \citep{Welsh2012}; Kepler-38b \citep{Orosz2012a}; Kepler-47b, 47c, \citep{Orosz2012b} and 47d \citep{Orosz2019}; PH1b/Kepler-64b \citep{Kostov2013,Schwamb2013}; Kepler-413b \citep{Kostov2014}; Kepler-453b \citep{Welsh2015}; and Kepler-1647b \citep{Kostov2016}. The most recently discovered CBP, Kepler 1647b, is a Jupiter-sized planet orbiting its 11-day host eclipsing binary every $\sim$1108 days in the habitable zone \citep{Kostov2016}.


There are several interesting architectural properties for the Kepler CBPs. First of all, there is a prevalence of small mutual inclinations between planetary orbits and host binary orbits, which is not due to selection effects, and this suggests that the occurrence rate of circumbinary systems is similar to that of single-star systems \citep{Armstrong2014,Martin2014,Li2016}. The results of BEBOP radial velocity survey tend to agree with such coplanarity, suggesting there isn't a large population of massive misaligned CBPs \citep{Martin2019}. 
Moreover, Kepler CBPs preferentially orbit around binaries with relatively long stellar orbital periods ($\gtrsim 7$ days). Several mechanisms have been proposed to explain this, such as Lidov-Kozai oscillations, which either drive planets in such systems out of transit configurations or remove the planets via instability \citep{Munoz15,Martin15,Hamers16};  
as well as stellar tidal effects, which cause the expansion of stellar orbits and lead to the ejection of closer-in planets \citep{Fleming2018}; UV photoevaporation and evection resonance were also brought up to explain such phenomenon \citep{Sanz-Forcada2014, Xu2016}. Lastly, most innermost Kepler CBPs reside close to the respective stability limits, except for Kepler-1647b. In-situ formation of giant planets at such close distances is a problem due to difficulties in planetesimal growth, which suggests that inward migration takes place in the formation pathway \citep{Paardekooper2012, Kley2014, Bromley2015, Silsbee2015}. 


There is likely a dominance of migration if the observed proximity of planets near the stability boundary is not due to selection effects, since planetary migration is an important process in the formation of CBPs near the stability limit. \citet{Li2016} studied the semi-major axis distribution of the innermost Kepler CBPs and found that the strength of evidence for a pile-up largely depends on whether Kepler-1647b is representative of the current population or belongs to another population of CBPs. Therefore, given a currently limited sample of transiting CBPs, it is important to know whether Kepler-1647b is really as special as it seems, or is instead just a temporary incomplete picture of the known CBP population. 


\citet{Quarles2018} considered another equal-mass planet closer to stability limit on a coplanar orbit for all Kepler CBP systems, and found nearly half of them could host the additional planet. In this work, we specifically consider the possibility that there is a yet undetected inner planet near the stability limit in the Kepler-1647 binary system. Differences with the previous study are to identify the effects of an additional planet on the observations (transit, TTV, ETV) and allow for non-coplanar orbits of the inner planet. The hypothetical planet needs to satisfy several conditions simultaneously, including: residing in a stable region, not transiting the host binary during Kepler mission lifetime, and having a small gravitational perturbation on other bodies in the system. 

This article is organized as follows. In section \ref{sec:sl}, we study orbital stability in the Kepler-1647 system, which is a basic requirement for a possible inner planet. In section \ref{sec:tp}, we investigate the transit probability of the inner planet. In section \ref{sec:ttv}, we explore the transit timing variations (TTVs) of the outer planet and eclipse timing variations (ETVs) induced by gravitational perturbations of the inner planet. And in section \ref{sec:ttvtp}, we combine the transit probability, TTV and ETV to better constrain the possible inner planet.

\section{Stability Limit} \label{sec:sl}

For an inner planet to exist, the minimum requirement is that such a body is on a stable orbit. Dynamical stability in circumbinary systems has been previously studied extensively \citep{Dvorak1989,Holman1999,Pilat-Lohinger2003,Musielak2005,Doolin2011,Kratter2014,Quarles2018}. In particular, \citet{Holman1999} numerically derived the stability limit of planets, assuming them to behave like test particles initially on circular, coplanar orbits, as a function of the host binary mass ratio and binary eccentricity. \citet{Doolin2011} identified how the potential stability changes when the planetary mutual inclination varies from 0$^\circ$ to 180$^\circ$. 

To give an intuitive picture of the Kepler-1647 system, the 11-day period central binary consists of stars with 1.22 and 0.97 M$_{\odot}$ on slightly eccentric orbits ($e=0.16$). The 1.5 M$_{J}$ planet resides on an 1108-day orbit with $e=0.06$. The binary and planetary orbital planes are closely aligned ($i\approx3^{\circ}$).

First, we numerically investigate the dynamical stability in the Kepler-1647 system because previous investigations \citep{Holman1999,Quarles2018} largely used general initial states of the binary system and considered only initially coplanar orbits, where here we calculate the stability limit following the derived parameters in \cite{Kostov2016} and allow the planetary orbit to be inclined. We performed a large group of N-body simulations using a modified version of MERCURY6 that includes a module designed for circumbinary systems \citep{Chambers2002}. 

In these simulations, we fix the initial parameters of the binary and the outer planet according to the best-fit result from observations \citep[see Table 5 in][]{Kostov2016}. We add to the system an additional Earth-mass inner planet that begins on a circular orbit, where we varied its semi-major axis (from $1.2a_{bin}$ to $4a_{bin}$ with $0.05a_{bin}$ steps) and mutual inclination to the binary orbital plane (from $0^\circ$ to $180^\circ$ with $3^\circ$ steps). The initial longitude of ascending node and  mean anomaly are chosen randomly from a uniform distribution ($0^\circ-360^\circ$). Each choice of initial semi-major axis and mutual inclination included 40 simulations. The inner planet is deemed stable if it did not undergo a collision with other three bodies in the system (inner binary or outer planet) nor did the radial distance to the test planet exceed 10 AU within the $10^{5}$ years of simulation time. Note that we include the outer planet Kepler 1647b in our simulations for completeness, but Kepler 1647b does not affect the stability limit, because it is very far away from the stellar binary ($\sim 21.3a_{bin}$). In addition, we note that starting the inner planet at its forced eccentricity does not reduce the stability limit. This is because the secular forced eccentricity cannot be generalized near the stability limit, where the non-secular effects (e.g., mean motion resonances) dominate, and thus, the maximum eccentricity that can be reached by the inner planet even starting at the forced eccentricity is similar to that starting with a circular orbit.

\begin{figure}
\plotone{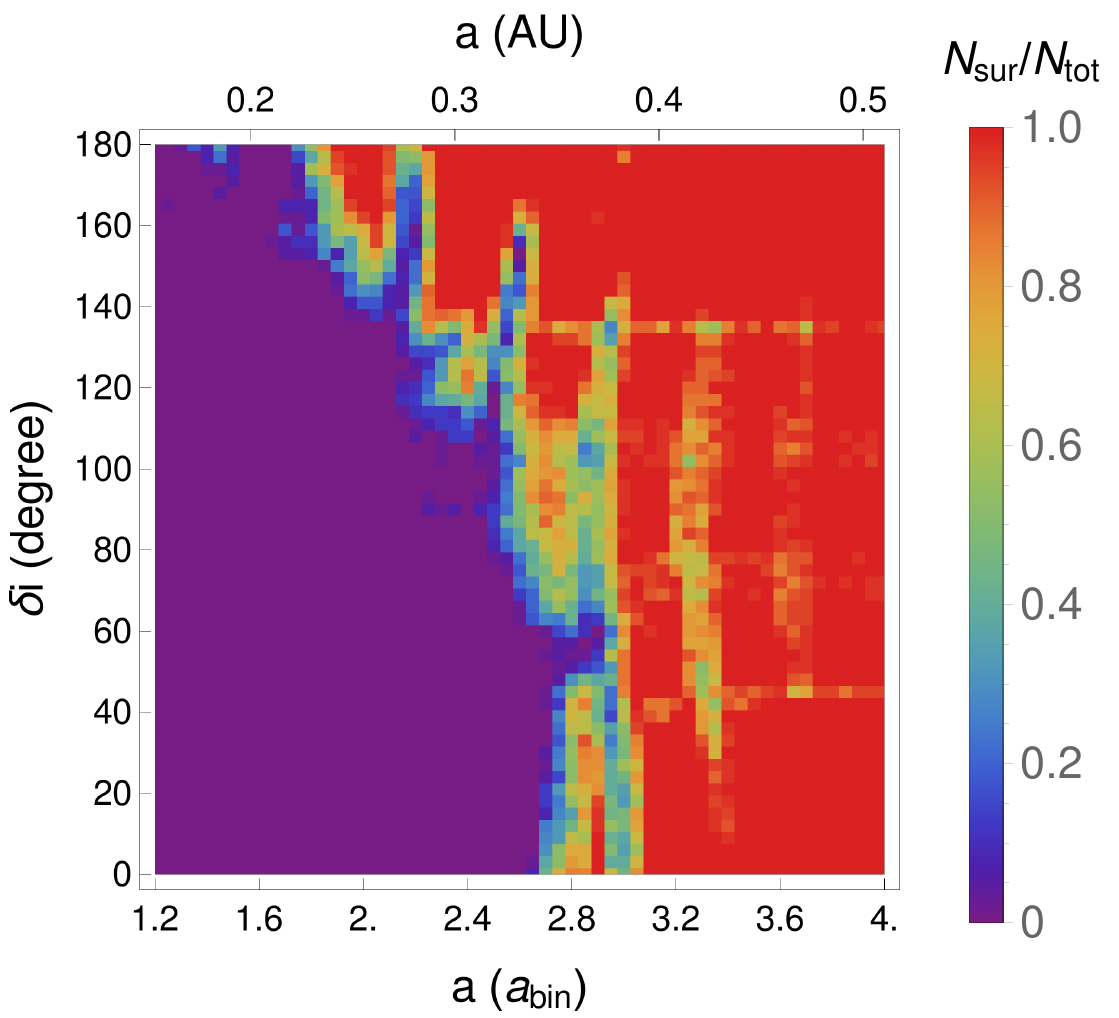}
\caption{Probability that a yet undetected inner planet in the Kepler-1647 binary system can survive ejection (beyond 10 AU) or collision with the inner binary, as a function of the initial planetary semi-major axis and mutual inclination using numerical simulations for $10^{5}$ years. Retrograde orbits are generally more stable than prograde orbits.  \label{fig:1}}
\end{figure}

Figure \ref{fig:1} shows the result of the simulations, where the color code denotes the fraction of surviving initial conditions for each trial of planetary semi-major axis and mutual inclination. Red cells denote when all forty trials of the simulated inner planets survived, while the purple cells designate wholly unstable initial conditions. We can see that retrograde orbits ($\delta i > 90^\circ$) are generally more stable than prograde orbits ($\delta i \leq 90^\circ$), consistent with previous studies \citep{Doolin2011,Li2016}, though planets on such orbits are presumed rare. In the low mutual inclination region ($\delta i \lesssim 20^\circ$), there is a stability island at 2.85 to 2.90 $a_{bin}$. We ran additional simulations for coplanar cases with 0.01 $a_{bin}$ intervals to find a more accurate semi-major axis value for the stability island. For each semi-major axis value, we ran 90 simulations varying the initial mean anomaly from $0^\circ$ to $356^\circ$ with $4^\circ$ steps. The results show that the stability island extends from 2.88 to 2.90 $a_{bin}$ for the low inclination near coplanar configurations. When the inclination is higher, there are vertical features with decreased survivability fractions at $\sim$3.3 $a_{bin}$. These features correspond to the 6:1 and 7:1 mean motion resonances with the binary \citep[e.g.,][]{Doolin2011,Quarles2016}.

In this paper, we define the stability limit by a critical semi-major axis, which is the shortest semi-major axis where the test particles at any chosen longitude survive the full integration time. Therefore, the critical semi-major axis ($a_{crit}$) is 2.88 $a_{bin}$ for a near coplanar inner planet, which is consistent with the results based on the chaos indicator ``Mean Exponential Growth factor of Nearby Orbits'' (MEGNO) discussed in \citet{Kostov2016}.

\begin{figure}
\plottwo{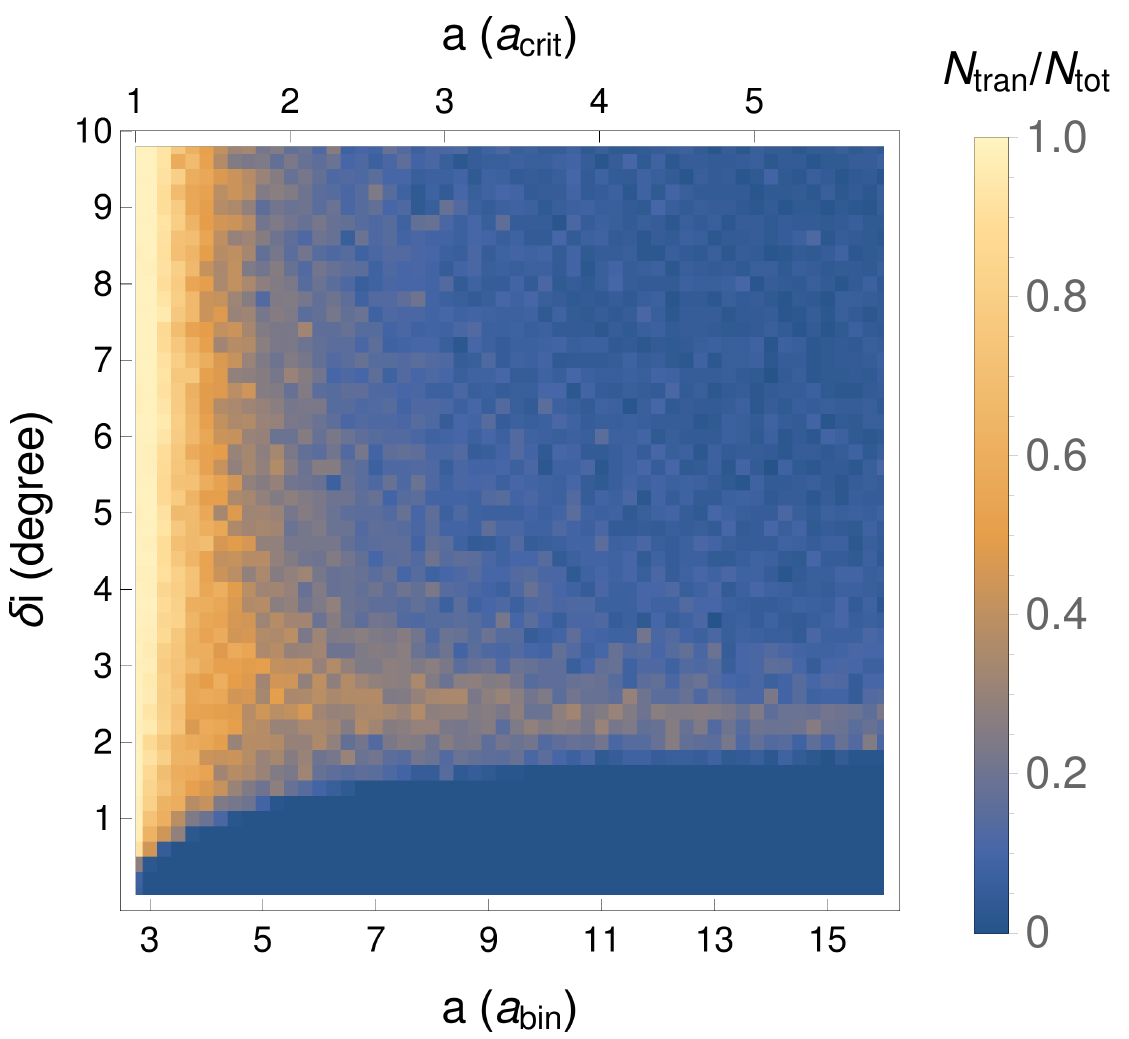}{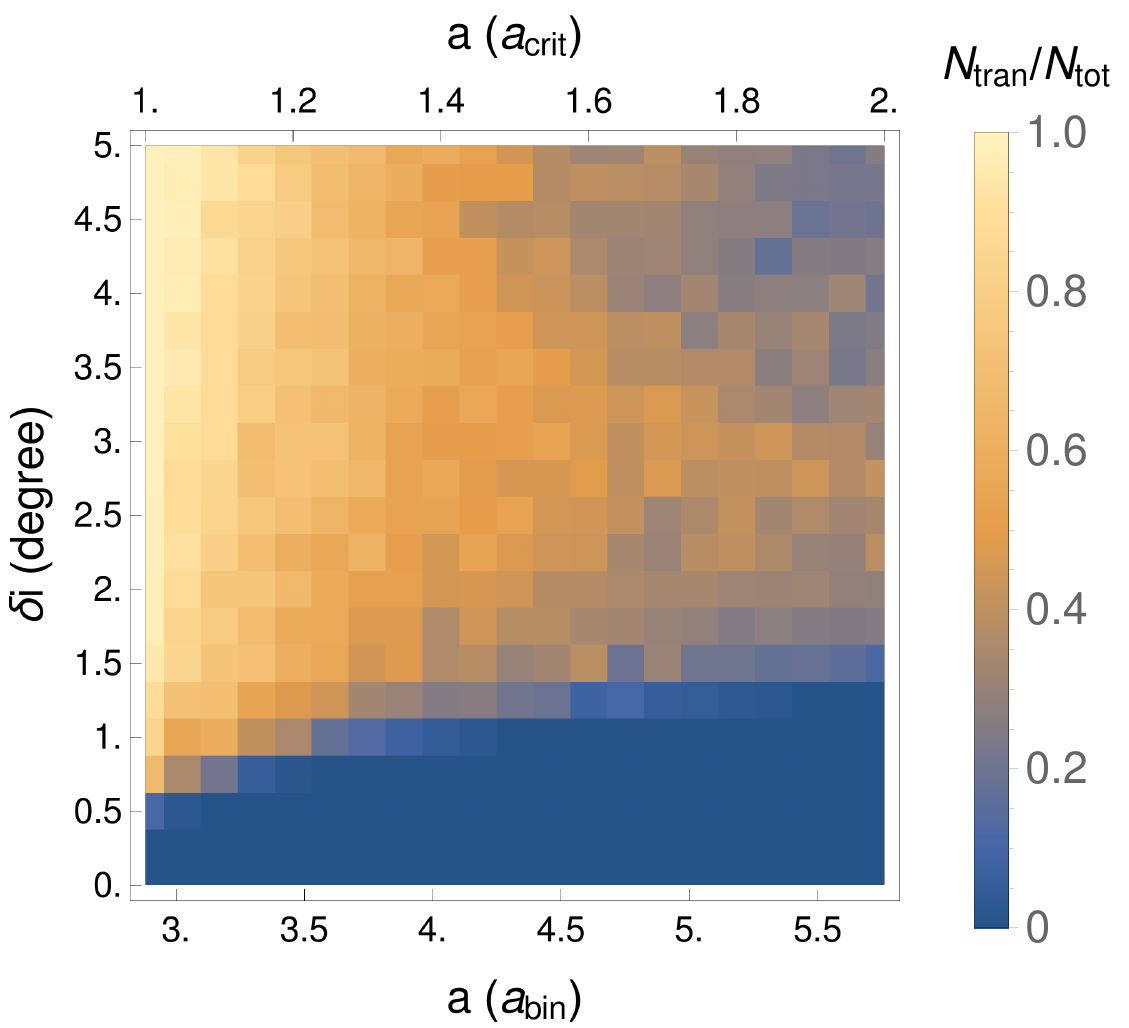}
\caption{Transit probability for an inner planet in the Kepler-1647 system during the 4-year Kepler mission lifetime considering a range of semi-major axis and mutual inclination values. The inner planet would not transit at all if the mutual inclination is lower than a critical inclination ($\delta i(a) \lesssim 2^\circ$).  The panel on the right shows a zoomed-in view considering the range of values ($a < 2a_{crit}$ and $\delta i \lesssim 4^\circ$) common among the known transiting CBPs.\label{fig:3}}
\end{figure}

\section{Transit Probability of the Inner Planet Near the Stability Limit} \label{sec:tp}
\citet{Kostov2016} performed a thorough visual inspection of the light curve of the Kepler-1647 system, and they didn't find any additional transits. Therefore, the existence of an inner planet can be ruled out if it could have transited during the Kepler mission lifetime. We consider the Kepler mission lifetime as a continuous 4-year period and use the transit probability to study this constraint. Previous studies have explored the transit probability of CBPs both analytically and numerically \citep{Martin2014,Martin2015,Li2016,Martin2017}. For simplicity, we calculate the transit probability using N-body simulations. We used the Rebound N-body integrator, along with the Reboundx modules \citep{Rein2015}, to study the transit probability.
In Reboundx there are modules to include general relativistic precession and tidal apsidal precession which we use in all of our simulations. The initial parameters of the binary and the outer planet are prescribed using the observationally derived values where the zero time is BJD 2455000 following \citet{Kostov2016}. Within this convention the start time is -40 days and the simulation ends at 1400 days.  In addition to the binary stars and the outer Jovian planet, we place an Earth-mass planet on a circular orbit in the stable region between the binary and outer planet (see Section \ref{sec:sl}), varied its semi-major axis and mutual inclination relative to the binary plane, and chose the initial longitude of ascending node and mean anomaly from a uniform distribution ($0^\circ-360^\circ$).

Figure \ref{fig:3} shows the that a close-in planet ($a \sim 3a_{bin}$), would have transited within the Kepler mission lifetime as long as the mutual inclination is greater than $\sim$0.5$^\circ$. Simulations in the left panel cover a range of semi-major axis values within the stable region and mutual inclinations up to 10$^\circ$, where each cell includes 50 simulations. The panel on the right shows a zoomed in view with a smaller range of semi-major axis ($1-2a_{crit}$) and mutual inclination ($\delta i \leq 5^\circ$) for the inner planet, where each cell includes 200 simulations. $a_{crit}$ is obtained using our simulational results in section \ref{sec:sl}. This smaller range of parameters is informed by the parameters of the other known Kepler CBPs, which should have a higher chance than other regions in the left panel.  However, this is contingent upon whether the other known CBPs are representative of the broader population. 

We find a critical mutual inclination between $\sim0.5^\circ$ and $\sim1.5^\circ$, depending on the value of the semi-major axis, where the transit probability changes steeply between zero and a relatively high value. The zero probability region under this critical mutual inclination results from a roughly two degree angle between the binary orbital plane and our line of sight. Above the critical inclination, the transit probability is close to unity when the planet is near $a_{crit}$ due to a high orbital precession rate ($\sim$53$^{\circ}$/yr at $a_{crit}$), suggesting that transits are nearly inevitable in this region. The transit probability decreases as semi-major axis increases, signifying that planets with a larger semi-major axis have a higher chance of being missed in a photometric survey like the Kepler mission. As a result, an inner planet has higher chance to reside at a low mutual inclination ($\delta i \lesssim1.5^\circ$), or at a higher inclination ($\gtrsim3^\circ$), but with a larger semi-major axis. 

In our model we did not consider the transit detection limit, so the result can only be applied to planets with detectable mass/radius. We did a rough estimation on the mass/radius limits for detecting a planet in Kepler-1647 based on signal/noise level. A planet with a radius less than 3 Earth radii would have been undetectable in the Kepler data. Using the M--R relation in \citet{Chen2017}, we place a lower limit on the mass of $\sim 10$ Earth masses. Note that this is only a crude estimate and detailed analysis on the limiting mass involves injection/retrieval methods, which is beyond the scope of our work.

\begin{figure}
\plotone{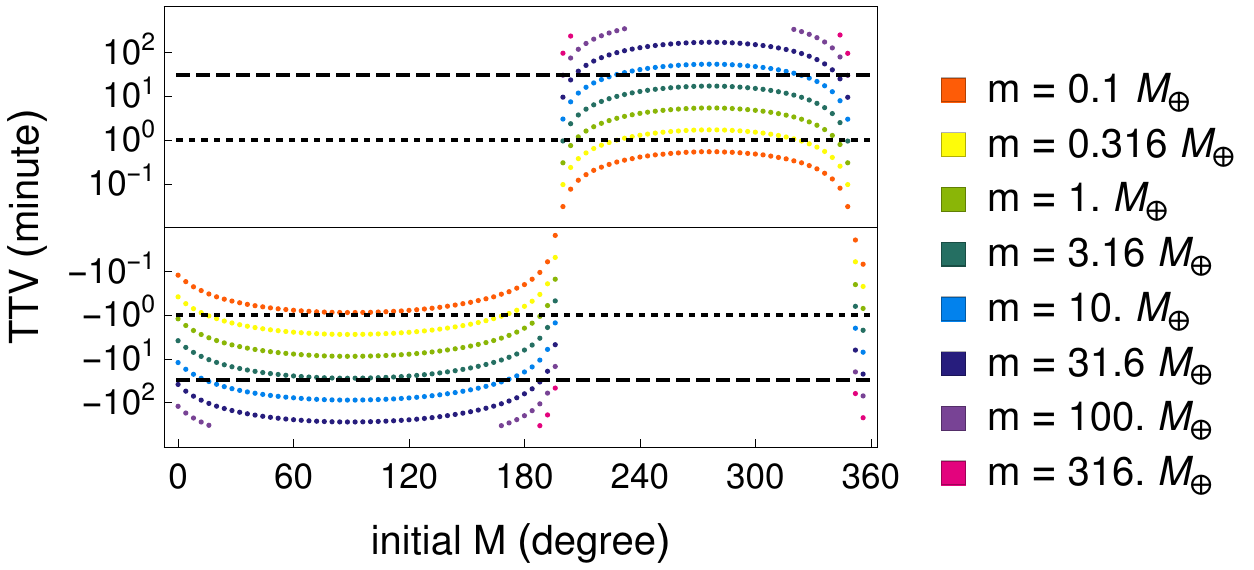}
\caption{Transit timing variations (TTVs) on the second observed transit as a function of the planetary initial mean anomaly induced by an inner planet mass ranging from a Mars to Jupiter analog. The black dashed lines represent 24 min observational uncertainty. Unsurprisingly, more massive planets lead to larger TTVs while a Mars-mass inner planet's TTVs (m = 0.1 M$_{\oplus}$) are below 1 min and a 30 M$_{\oplus}$ planet's TTVs are generally above the 24 min observational uncertainty.
\label{fig:4}}
\end{figure}

\begin{figure*}
\plotone{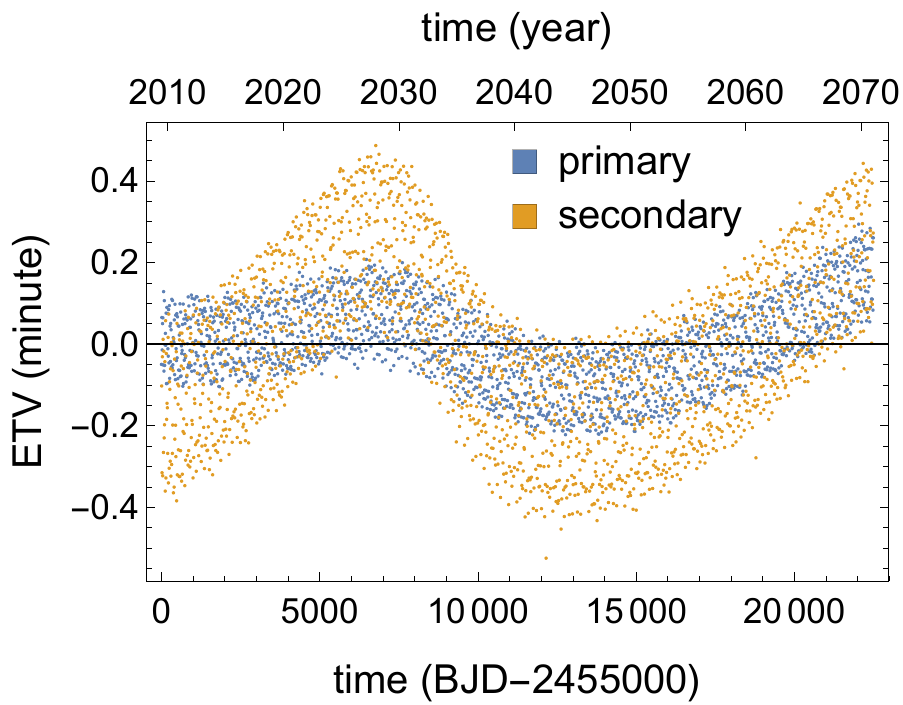}
\caption{
Eclipse timing variations (ETVs) of Kepler-1647 host binary induced by an inner planet with 100 M$_{\oplus}$, as a function of time since BJD 2455000. The blue and the orange dots represent ETVs for primary and secondary eclipse respectively. In this case, the inner planet begins with mean anomaly M = 135$^{\circ}$.
\label{fig:5}}
\end{figure*}

\begin{figure*}
\plottwo{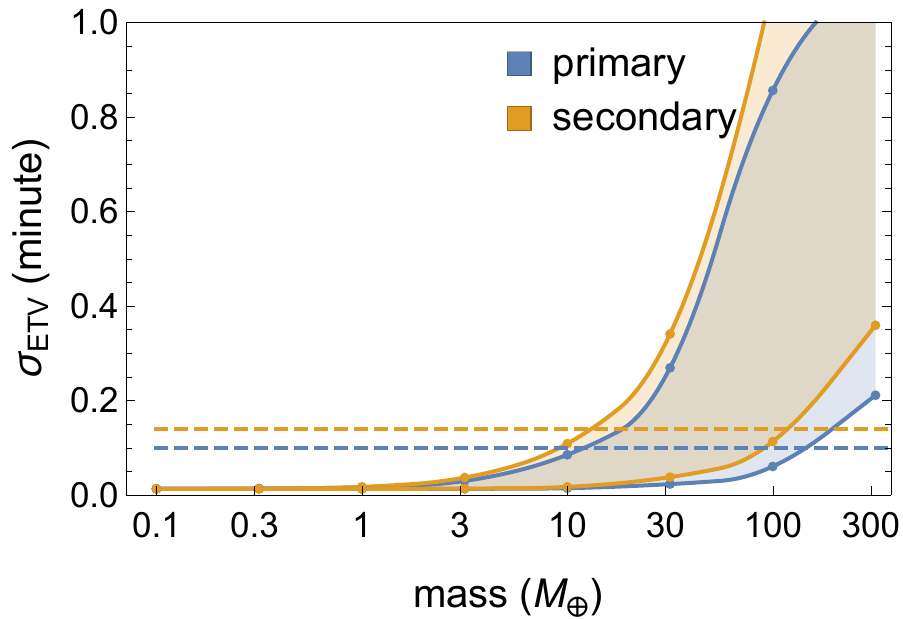}{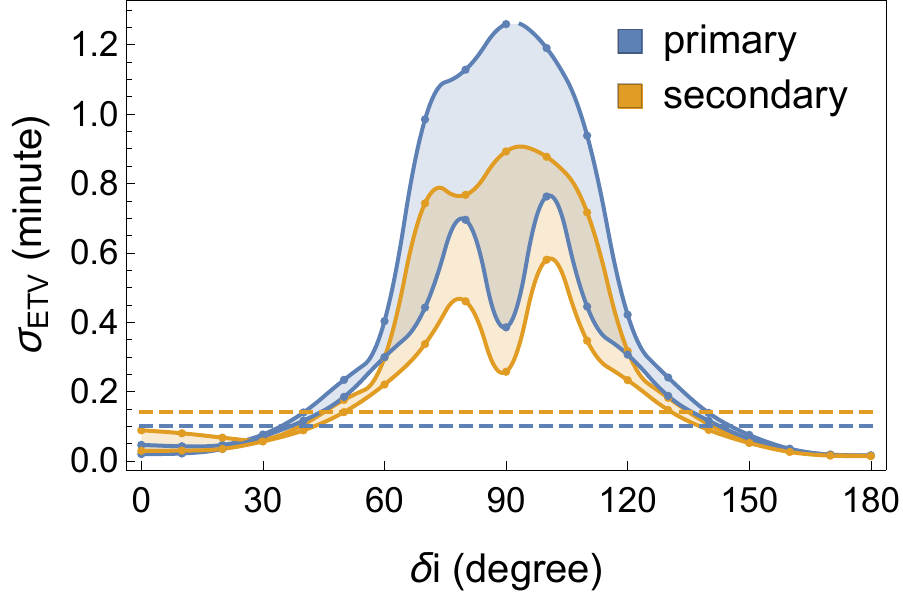}
\caption{
Left panel: standard deviations of ETVs as a function of planetary mass. Right panel: standard deviations of ETVs as a function of inner planet inclination relative to the stellar binary, where the planet mass is set to be 30 M$_\oplus$. In both panels, shaded areas indicate the possible range of the standard deviation, which come from different choices of initial mean anomaly, and dashed lines represent 1$\sigma$ observational uncertainties. The lower limits reflect the degree of scattering. An inner planet minimally requires a mass beyond $\sim 150$M$_\oplus$ to make the standard deviations of both the primary and secondary eclipses surpass the observational uncertainties regardless of initial mean anomaly. In addition, the standard deviation of the ETV increases when the mutual inclination is around $40-140^\circ$.
\label{fig:ETV}}
\end{figure*}

\section{TTVs and ETVs From an Inner Planet} \label{sec:ttv} 

The presence of an inner planet can affect the motion of the stellar binary and the known outer planet leading to discrepancies in the mid-eclipse (ETV) or mid-transit (TTV) times, respectively.  We find 
mid-transit and mid-eclipse times through photodynamical simulations using \texttt{photodynam} \citep{Carter2011,Pal2012}
. In these simulations, we begin the inner planet near the stability limit while varying its mass to estimate the TTVs and the ETVs.


Three transits of Kepler-1647b were detected during the 4-year Kepler mission. The observed mid-transit times are -3.0018, 1104.9510, 1109.2612 (in days, relative to 2455000[BJD]), transiting star B, star A, and star B, respectively. \cite{Kostov2016} reported 1$\sigma$ uncertainties of those measurements to be around five minutes. Using the initial conditions given in \cite{Kostov2016} as initial parameters, we calculated the numerical mid-transit times of the observed three transits to be -3.0035, 1104.9512, and 1109.2645, which are within 1$\sigma$ uncertainties of the observed values. Thus, we used our values from the simulations as a baseline for calculating the transit timing variations (TTVs) for the cases with an inner planet close to the stability limit. 

In these simulations, we used the same initial parameters for the binary and the outer planet, while placing an inner planet on an initially circular coplanar orbit as before. We varied the mass, semi-major axis, and initial mean anomaly of the inner planet. From our first test simulations, we found that the dependence of TTVs on the semi-major axis are much weaker than that on the planetary mass and initial mean anomaly, especially near $a_{crit}$. Therefore in our more detailed simulations, we placed the inner planet at $a_{crit}$, and varied its mass from 0.1--316M$_{\oplus}$ ($\sim$Mars -- Jupiter mass) geometrically, with $10^{0.5}$ common ratio, and the initial mean anomaly of the inner planet from $0-360^\circ$. 

Figure \ref{fig:4} shows the calculated TTV on a logarithmic scale (in minutes) induced on the transit of the outer planet across star A ($\sim$1104.9512 days). The TTVs increase approximately linearly with the inner planetary mass. For a given mass, the most extreme values of TTV appear when the initial mean anomaly is near 90$^\circ$ and 270$^\circ$. If we assume the orbits of known bodies are measured precise enough, we can conservatively rule out those masses producing a TTV $\gtrsim 24$ min (black dashed lines), which is the uncertainty of orbital period of Kepler-1647 b. This places an upper limit on the planetary mass to roughly 30 M$_{\oplus}$. Moreover, A Saturn-mass planet can produce a TTV $\gtrsim 2$ h, 
while a Mars-mass planet produces a TTV under 1 min through one orbit. 

Since there is only one full orbit of Kepler-1647b observed, the ``TTV'' can be due to wrong estimate of Kepler-1647b orbital parameters, which did not consider the effects of an inner planet. To better understand whether there could be an inner planet based on the TTV measurements, we provide photodynamical TTV results after ten orbits ($\sim$30 years) from the first observed transit for future reference. We used the same initial parameters for the binary and the outer planet, and placed an inner planet on an initially circular coplanar orbit at $a_{aver}=1.35126\:a_{crit}$ (which is the average semi-major axis, in units of $a_{bin}$, of the other eight innermost Kepler CBPs), choosing an initial mean anomaly at 270$^\circ$. The planetary masses were varied from 0.1--31.6M$_{\oplus}$ geometrically with $10^{0.5}$ common ratio. Table \ref{tab:1} shows the numerical results of our simulations, where the upper table and the lower tables are for transits across star A and star B, respectively. The first and eighth rows are the mid-transit times (in days) without an inner planet, relative to BJD 2455000. In some orbits there are multiple transits across a star separated by a few days, and we only include the first one in each orbit. The remaining rows are the calculated TTVs (in minutes) on the outer planet induced by an inner planet. Notice that the units in the first row and the remaining rows are different. The dashes ($-$) in the table denote those transits do not happen due to inner planet's influence. Underlined values in the table are the transits that overlap with eclipses, namely syzygies. The next pair of transits of the outer planet across star A and star B are expected to happen around 2021-Jul-30 and 2021-Jul-26 respectively (bold column). Observations at that time would greatly enhance our knowledge of the system because a 3--30 Earth-mass inner planet can produce a TTV on the order of hours. In particular, the next three pairs will happen in 2021, 2024 and 2027, and CHEOPS \citep{Broeg2013} may be able to detect the TTVs since it is scheduled to launch at the end of this year (2019). TESS after the primary mission \citep[e.g.,][]{Kostov2016} and PLATO 2.0 \citep{Rauer14} may also detect the future transits of Kepler 1647b. 


We also investigated eclipse timing variations (ETVs) due to the inner planet. In our simulations, we placed an inner planet on an initially circular coplanar orbit at $a_{crit}$, while varying its mass and initial mean anomaly. Figure \ref{fig:5} shows one of the cases. In this simulation, the 100 M$_{\oplus}$ inner planet begins with mean anomaly $M = 135^{\circ}$, running for 1000 binary periods ($\sim$30 years). We found that the scattering of ETVs increases as the mass increases, while the long-term trends are affected by the initial mean anomaly. In particular, for some starting mean anomalies, the long-term trend has no variablity and the magnitudes of ETVs are only due to the internal scattering. 

To find out the exact relation between ETV and planetary mass, we varied the planet mass from 0.1 -- 316M$_{\oplus}$ and began the inner planet at  $M=0^\circ, 45^\circ, ..., 315^\circ$. We use standard deviation to quantify the magnitude of ETV.  The left panel of Figure \ref{fig:ETV} shows the ranges of standard deviations of ETVs, where the dashed lines represent 1$\sigma$ observational uncertainties. The lower limits of the standard deviations reflect the magnitudes of scattering. We can see from the figure that the secondary eclipses have larger standard deviations than the primary eclipses. As planetary mass increases beyond $\sim$150M$_{\oplus}$, the standard deviations of primary and secondary eclipses both surpass the observational uncertainties for independent of the choice in the starting mean anomaly. This suggests that a planetary mass beyond $\sim$150M$_{\oplus}$ could be considered highly unlikely by ETV. In contrast, from $\sim$10M$_{\oplus}$ to $\sim$150M$_{\oplus}$ , the standard deviations can only sometimes surpass the observational uncertainties due to the long-term trend. However, the scatter in the ETV is small, so for short-term observation, the standard deviations of ETVs would still be lower than the uncertainty values. Future measurements of the stellar eclipse times can help reduce the observational uncertainties and further constrain the planetary mass.

So far, we considered ETVs due to a near coplanar inner planet. To illustrate that our results are robust, we allow the planet to be inclined from the plane of the stellar binary. We set the planetary mass to be 30M$_\oplus$ for illustration, and we set the planetary semi-major axis to be 3.1$a_{bin}$, since the stability island in the near coplanar region does not extend to the high inclination configurations. The results are shown in the right panel of Figure \ref{fig:ETV}. Shaded areas indicate the possible range of the standard deviation, obtained based on 8 different choices of initial mean anomalies (0, 45, ..., 315 deg), and dashed lines represent 1$\sigma$ observational uncertainties. The ascending node is drawn from a uniform distribution.

The figure shows that the standard deviation of the ETV increases when the inclination relative to the stellar binary is larger, between $\sim$40$^{\circ}$ and $\sim$140$^{\circ}$. In this region, the minimum ETVs exceed the 1$\sigma$ observational uncertainties, indicating that Kepler-1647 is unlikely to host an inner planet with a mass higher than 30M$_\odot$ in an orbit with large inclination ($40-140^\circ$). When the inclination increases above $\sim 140^\circ$, the standard deviation decreases and converges to a similar ETV magnitude as the near coplanar case.
Note that EDVs (eclipse duration variations) unfortunately do not provide extra useful information, due to the large observational uncertainties in the ingress and egress of the eclipses.

\begin{table}[h!]
\caption{Transit times and TTVs of Kepler-1647b for ten orbits after the first observed transit.
\label{tab:1}}
\begin{tabular}{*{1}{*{12}{l}}}
\hline
\hline
$\text{No inner [day]}$ & $0.37$ & $\underline{1104.96}$ & $2209.55$ & $3314.55$ & $\textbf{4425.74}$ & $5530.30$ & $6635.01$ & $7746.50$ & $\underline{8851.08}$ & $9955.68$ & $11060.72$ \\
$0.1M_{\oplus} [min]$ & $0$ & $\underline{0}$ & $1$ & $3$ & $\textbf{2}$ & $2$ & $4$ & $3$ & $\underline{4}$ & $5$ & $10$ \\
$0.3M_{\oplus} [min]$ & $0$ & $\underline{1}$ & $3$ & $9$ & $\textbf{6}$ & $8$ & $12$ & $11$ & $\underline{12}$ & $15$ & $31$ \\
$1.0M_{\oplus} [min]$ & $0$ & $\underline{5}$ & $10$ & $28$ & $\textbf{19}$ & $24$ & $37$ & $36$ & $\underline{37}$ & $47$ & $99$ \\
$3.1M_{\oplus} [min]$ & $0$ & $\underline{15}$ & $33$ & $88$ & $\textbf{60}$ & $76$ & $119$ & $112$ & $\underline{118}$ & $151$ & $333$ \\
$10.0M_{\oplus} [min]$ & $0$ & $\underline{47}$ & $105$ & $293$ & $\textbf{\underline{189}}$ & $242$ & $388$ & $350$ & $\underline{373}$ & $488$ & $1482$ \\
$31.6M_{\oplus} [min]$ & $0$ & $\underline{148}$ & $336$ & $1136$ & $\textbf{\underline{593}}$ & $783$ & $1415$ & $\underline{1074}$ & $1203$ & $1727$ & $-$ \\
\end{tabular}
\begin{tabular}{*{1}{*{12}{l}}}
\hline
\hline
$\text{No inner [day]}$ & $-3.00$ & $1109.27$ & $2213.73$ & $\underline{3317.95}$ & $\textbf{4422.21}$ & $5534.62$ & $6638.90$ & $7743.11$ & $8847.56$ & $9959.84$ & $\underline{11064.06}$ \\
$0.1M_{\oplus} [min]$ & $0$ & $1$ & $1$ & $\underline{1}$ & $\textbf{2}$ & $2$ & $2$ & $2$ & $5$ & $3$ & $\underline{3}$ \\
$0.3M_{\oplus} [min]$ & $0$ & $2$ & $2$ & $\underline{3}$ & $\textbf{5}$ & $7$ & $6$ & $8$ & $15$ & $10$ & $\underline{10}$ \\
$1.0M_{\oplus} [min]$ & $0$ & $6$ & $7$ & $\underline{10}$ & $\textbf{16}$ & $21$ & $20$ & $24$ & $49$ & $32$ & $\underline{33}$ \\
$3.1M_{\oplus} [min]$ & $0$ & $18$ & $23$ & $\underline{31}$ & $\textbf{52}$ & $65$ & $\underline{63}$ & $78$ & $159$ & $101$ & $\underline{105}$ \\
$10.0M_{\oplus} [min]$ & $0$ & $55$ & $72$ & $\underline{99}$ & $\textbf{167}$ & $202$ & $\underline{199}$ & $250$ & $-$ & $314$ & $333$ \\
$31.6M_{\oplus} [min]$ & $0$ & $171$ & $224$ & $315$ & $\textbf{567}$ & $612$ & $\underline{626}$ & $847$ & $-$ & $\underline{963}$ & $1109$ \\
\hline
\hline
\end{tabular}
\tablecomments{The upper table and the lower table are for transits across star A and star B respectively. The fist row of each table is the mid-transit times (in days) without an inner planet, relative to BJD 2455000. The remaining rows are the calculated TTVs (in minutes) induced by an inner planet with different mass. Bold column is the next pair of transits in July 2021. Underlined elements are syzygies. The dashes ($-$) denote when the transits do not happen due to inner planet's influence. }
\end{table}

\begin{figure}[ht]
\plotone{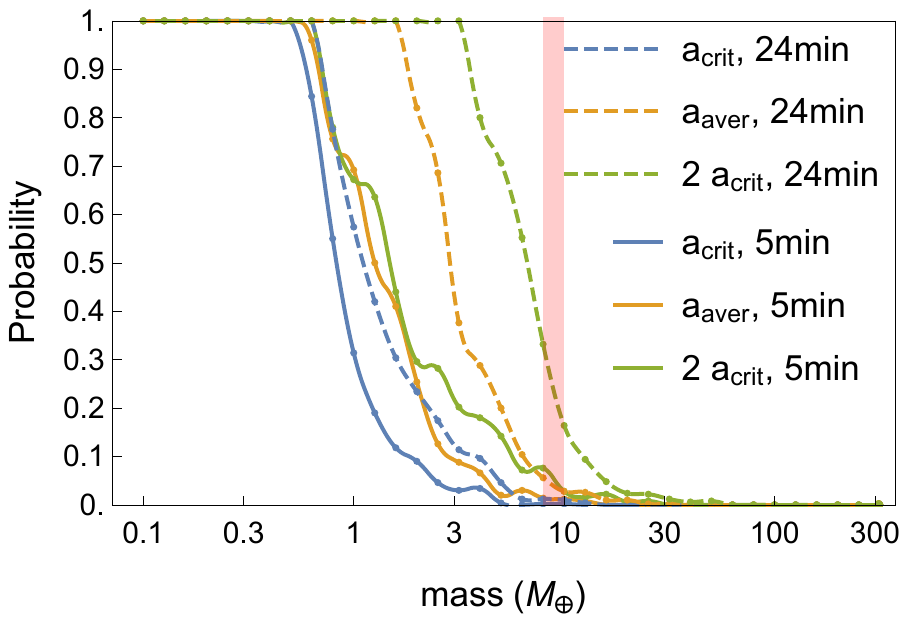}
\caption{Probability for an inner planet to escape detection considering a range of planetary masses and semi-major axis values. Assuming the inner planet would not have been detected if the induced TTV caused to the outer planet is $<$5 min (solid lines) or $<$24 min (dashed lines), the induced accumulated ETV to be lower than 0.30 min for the primary and 0.42 for the secondary around the end of Kepler mission (3$\sigma$ uncertainties) and for those having a mass $\gtrsim$10M$_{\oplus}$ (right to the red line), the planet did not transit during Kepler mission lifetime. We can see planets $\gtrsim$30M$_\oplus$ have a very low probability to escape detection and can be ruled out. \label{fig:6}}
\end{figure}

\section{Combining TTVs and ETVs with the Transit Probability} \label{sec:ttvtp}

In Section \ref{sec:tp} and Section \ref{sec:ttv}, we explored the transit probability and possibility of TTVs and ETVs independently. However, they can be considered simultaneously to make a comprehensive constraint, as we will elaborate in this section. We define the combined probability for an undetected planet to include the probability that an inner planet didn't transit during Kepler four--year mission (if its size lies above the transit detection limit), meanwhile only causing small TTV and ETV on the outer planet and the host binary. In addition, we consider the differences between eclipse times of the 4-body model (with an inner planet) and that of the 3-body model (without an inner planet) as constraints due to ETVs. In this way we find a stronger constraint, as we already have observed eclipses times during the Kepler mission. 

For this calculation an inner planet is added to the system and begins on a circular, near-coplanar orbit close to the stability limit. The mutual inclination relative to the binary orbital plane was drawn from a normal distribution with a 0$^\circ$ mean value and 3$^\circ$ standard deviation, since CBPs are near co-planar with stellar binary orbits. The inner planet's mass was varied from 0.1--316M$_{\oplus}$ ($\sim$Mars -- Jupiter mass) geometrically, with $10^{0.1}$ common ratio. The semi-major axis was chosen to be $a_{crit}$, $a_{aver}$, and $2 a_{crit}$. TTV requirements were placed on the outer planet's transit across the primary star at under 24 min or 5 min. We also require that the accumulated ETV of the 4-body model (with an inner planet) to be below 0.30 min for primary and 0.42 min for secondary compared with the 3-body model (without an inner planet) during the four years of the observational window, which are based on the 3$\sigma$ uncertainties from CPOC residuals in \citet{Kostov2016}. And for a planet with mass larger than 10 M$_{\oplus}$, requirement that the planet did not transit were also placed. For each chosen planetary mass, we have 500 simulations with the inner planet starting with a random longitude of ascending node and a random mean anomaly both drawn from a uniform distribution ($0^\circ-360^\circ$). Based on the ETV results alone, we find that if the planet is at $2a_{crit}$, all of the 500 runs produce ETVs above 3$\sigma$ when the planetary mass is above $100$M$_{\oplus}$. This suggests that we can rule out any inner planets with masses $\gtrsim 100$M$_{\odot}$ within twice the instability limit based on the four years of stellar eclipsing data. Similarly, if the planet is within $a_{aver}$, we can rule out planets with masses above $30$M$_{\odot}$ based on the four years of ETV results alone.



The solid lines and the dashed lines, in Figure \ref{fig:6}, represent the TTV requirement for 5 min and 24 min, respectively. The red vertical line shows the transit detection limit at $\sim$10M$_{\oplus}$. In the left part, the probability is only constrained by TTV and ETV because a planet with such size is too small to be captured by transit, as discussed in Section \ref{sec:tp}
. 
On the right part, the probability is constrained by TTV, ETV and transit detection. 
For the TTV $<$ 5 min curves, the possibilities are depend mainly on the TTV and the possibilities are constrained by ETV using the TTV $<$ 24 min curves. We can see from the former that the possibilities of hiding reach small values at several Earth masses for all three semi-major axis values chosen. As for the latter, the possibility of a $\sim$5M$_{\oplus}$ inner planet is small near $a_{crit}$ and that of a $\sim$30M$_{\oplus}$ inner planet is extremely small even at $2a_{crit}$. 

The existence of an inner planet perturbs the orbit of the outer planet, Kepler-1647b, and it is possible that the observed orbital parameters, without taking into account the effects of an inner planet, is not accurate. Thus, we allowed the orbital parameters of Kepler-1647b to vary and evaluated the photometric light curves for Kepler-1647 using an eclipsing light curve (ELC) code \citep{Orosz2000,Welsh2015} that performs parameter estimation based upon a Differential Evolution Monte Carlo Markov Chain (DE-MCMC) routine \citep{Braak2006}.  Through our code, we reconfirm the results from \cite{Kostov2016} assuming there's no inner planet, and we performed a series of tests considering an additional planet near the stability limit.  We tested 5 different initial values for the putative inner planet (10, 30, 50, 100, \& 300 M$_\oplus$), where each test randomly varies the initial mutual inclination by 1$^\circ$ from the binary plane so that it does not cross the line of sight. Our tests revealed that the parameters of the inner stellar binary and Kepler-1647b changed very little (within the 3 body model uncertainties), because the inner planet is small and far away from Kepler-1647b. This reconfirmed that our analysis above applying the observed orbital parameters of Kepler-1647b following \cite{Kostov2016} is robust. Although we started the planet at the stability limit ($P_{c}$ $\sim$55 days), the DE-MCMC routine searches models with orbital periods from 50 -- 150 days.  Our best fit orbital periods for the inner planet settled $\sim$140 days ($>2\times$ $P_c$) when the initial value was equal to 50 M$_\oplus$ or more. This indicates that Kepler-1647 is unlikely to host an inner planet of masses above $\sim 50$M$_\oplus$ near the stability limit, consistent with our analysis above.

\section{Conclusion and Discussion} \label{sec:cd}

In this paper we studied the Kepler-1647 system, investigating the parameter space of a possibly undetected inner planet near the stability limit. First of all, we explored the dynamical stability of the system, particularly the dependence on semi-major axis and mutual inclination. We find the critical semi-major axis, $a_{crit}$, at $2.88a_{bin}$. 

Secondly, we studied the transit probability of the inner planet, where we uncovered a zero-probability zone under a critical mutual inclination, which is an ideal region for the planet to escape detection from photometric observations in the Kepler mission. The inner planet's gravitational influence can affect the eclipses of the inner binary or the transits of the known planet. We used mid-eclipse/transit times to determine the magnitude of the possible variations. We calculated the magnitude of the TTVs of the second observed transit as a function of the initial mean anomaly and planetary mass for an inner planet at $a_{crit}$. Typically, a 30 M$_{\oplus}$ planet would produce a TTV $\gtrsim 24$ min uncertainty. We also provide future TTV observation windows, which can be applied to future observations and will further constrain possible undetected planets with lower masses near the stability limit. In addition, we calculated the ETVs of the stellar binary including an inner planet near the stability limit. We found that planetary masses above $\sim 30$M$_{\odot}$ always produces an ETV above the observational uncertainties if the planet is within $a_{aver}$, the average distance of the observed innermost circumbinary planets scaled with the stability limit. By considering the constraints from transit probability (i.e., combining TTV and ETV constraints), we calculated the possibility for an inner planet to escape detection during the Kepler mission, we find an inner planet with mass $\gtrsim$30M$_{\oplus}$ can be ruled out within 2$a_{crit}$. 

It is possible that Kepler-1647b's orbit is different from the observational results, which assumed that there were no inner planets. We allowed the orbital parameters of the stellar binary and Kepler-1647b to vary and performed parameter estimation based upon a Differential Evolution Monte Carlo Markov Chain (DE-MCMC) routine \citep{Braak2006}, based on the photometric light curves for Kepler-1647 using an eclipsing light curve (ELC) code \citep{Orosz2000,Welsh2015}. We find we can rule out inner planet mass beyond around $\sim 30$M$_{\oplus}$ near the stability limit, consistent with our analysis based on TTV, ETV and transit probability of the inner planet.

There are several possible scenarios for Kepler-1647 system. Firstly, no inner planets exist near the stability limit. In this case, 
Kepler-1647 is indeed from a different population comparing with the other CBP systems. It shows that planetary migration close to the instability limit is not universal in CBP systems. Secondly, there is a low-mass inner planet near the stability limit. In this case, the evidence for a pile-up among CBPs would then be more significant. A dominance of inward migration would be supported if the environment near the stability limit does not favor in-situ formation even for low-mass planets. Any of these cases would make Kepler-1647 a interesting system to study and provide valuable information for theories concerning the formation of CBPs. 

\section*{Acknowledgments}
The authors would like to thank Veselin Kostov, David Martin and Dan Fabrycky for helpful discussions.





\bibliography{reference.bib}

\begin{thebibliography}{}
\expandafter\ifx\csname natexlab\endcsname\relax\def\natexlab#1{#1}\fi
\providecommand{\url}[1]{\href{#1}{#1}}

\bibitem[{{Armstrong} {et~al.}(2014){Armstrong}, {Osborn}, {Brown}, {Faedi},
  {G{\'o}mez Maqueo Chew}, {Martin}, {Pollacco}, \& {Udry}}]{Armstrong2014}
{Armstrong}, D.~J., {Osborn}, H.~P., {Brown}, D.~J.~A., {et~al.} 2014, \mnras,
  444, 1873

\bibitem[{{Beuermann} {et~al.}(2010){Beuermann}, {Hessman}, {Dreizler},
  {Marsh}, {Parsons}, {Winget}, {Miller}, {Schreiber}, {Kley}, {Dhillon},
  {Littlefair}, {Copperwheat}, \& {Hermes}}]{Beuermann2010}
{Beuermann}, K., {Hessman}, F.~V., {Dreizler}, S., {et~al.} 2010, \aap, 521,
  L60

\bibitem[{{Broeg} {et~al.}(2013){Broeg}, {Fortier}, {Ehrenreich}, {Alibert},
  {Baumjohann}, {Benz}, {Deleuil}, {Gillon}, {Ivanov}, {Liseau}, {Meyer},
  {Oloffson}, {Pagano}, {Piotto}, {Pollacco}, {Queloz}, {Ragazzoni}, {Renotte},
  {Steller}, \& {Thomas}}]{Broeg2013}
{Broeg}, C., {Fortier}, A., {Ehrenreich}, D., {et~al.} 2013, in European
  Physical Journal Web of Conferences, Vol.~47, European Physical Journal Web
  of Conferences, 03005

\bibitem[{{Bromley} \& {Kenyon}(2015)}]{Bromley2015}
{Bromley}, B.~C., \& {Kenyon}, S.~J. 2015, \apj, 806, 98

\bibitem[{{Carter} {et~al.}(2011){Carter}, {Fabrycky}, {Ragozzine}, {Holman},
  {Quinn}, {Latham}, {Buchhave}, {Van Cleve}, {Cochran}, {Cote}, {Endl},
  {Ford}, {Haas}, {Jenkins}, {Koch}, {Li}, {Lissauer}, {MacQueen}, {Middour},
  {Orosz}, {Rowe}, {Steffen}, \& {Welsh}}]{Carter2011}
{Carter}, J.~A., {Fabrycky}, D.~C., {Ragozzine}, D., {et~al.} 2011, Science,
  331, 562

\bibitem[{{Chambers} {et~al.}(2002){Chambers}, {Quintana}, {Duncan}, \&
  {Lissauer}}]{Chambers2002}
{Chambers}, J.~E., {Quintana}, E.~V., {Duncan}, M.~J., \& {Lissauer}, J.~J.
  2002, \aj, 123, 2884

\bibitem[{{Chen} \& {Kipping}(2017)}]{Chen2017}
{Chen}, J., \& {Kipping}, D. 2017, \apj, 834, 17

\bibitem[{{Doolin} \& {Blundell}(2011)}]{Doolin2011}
{Doolin}, S., \& {Blundell}, K.~M. 2011, \mnras, 418, 2656

\bibitem[{Doyle {et~al.}(2011)Doyle, Carter, Fabrycky, Slawson, Howell, Winn,
  Orosz, Prˇsa, Welsh, Quinn, Latham, Torres, Buchhave, Marcy, Fortney,
  Shporer, Ford, Lissauer, Ragozzine, Rucker, Batalha, Jenkins, Borucki, Koch,
  Middour, Hall, McCauliff, Fanelli, Quintana, Holman, Caldwell, Still,
  Stefanik, Brown, Esquerdo, Tang, Furesz, Geary, Berlind, Calkins, Short,
  Steffen, Sasselov, Dunham, Cochran, Boss, Haas, Buzasi, \&
  Fischer}]{Doyle2011}
Doyle, L.~R., Carter, J.~A., Fabrycky, D.~C., {et~al.} 2011, Science, 333,
  1602.
\newblock \url{http://science.sciencemag.org/content/333/6049/1602}

\bibitem[{{Dvorak} {et~al.}(1989){Dvorak}, {Froeschle}, \&
  {Froeschle}}]{Dvorak1989}
{Dvorak}, R., {Froeschle}, C., \& {Froeschle}, C. 1989, \aap, 226, 335

\bibitem[{{Fleming} {et~al.}(2018){Fleming}, {Barnes}, {Graham}, {Luger}, \&
  {Quinn}}]{Fleming2018}
{Fleming}, D.~P., {Barnes}, R., {Graham}, D.~E., {Luger}, R., \& {Quinn}, T.~R.
  2018, \apj, 858, 86

\bibitem[{{Hamers} {et~al.}(2016){Hamers}, {Perets}, \& {Portegies
  Zwart}}]{Hamers16}
{Hamers}, A.~S., {Perets}, H.~B., \& {Portegies Zwart}, S.~F. 2016, \mnras,
  455, 3180

\bibitem[{Holman \& Wiegert(1999)}]{Holman1999}
Holman, M.~J., \& Wiegert, P.~A. 1999, The Astronomical Journal, 117, 621.
\newblock \url{http://stacks.iop.org/1538-3881/117/i=1/a=621}

\bibitem[{{Kley} \& {Haghighipour}(2014)}]{Kley2014}
{Kley}, W., \& {Haghighipour}, N. 2014, \aap, 564, A72

\bibitem[{Kostov {et~al.}(2013)Kostov, McCullough, Hinse, Tsvetanov, Hébrard,
  Díaz, Deleuil, \& Valenti}]{Kostov2013}
Kostov, V.~B., McCullough, P.~R., Hinse, T.~C., {et~al.} 2013, \apj, 770, 52.
\newblock \url{http://stacks.iop.org/0004-637X/770/i=1/a=52}

\bibitem[{Kostov {et~al.}(2014)Kostov, McCullough, Carter, Deleuil, Díaz,
  Fabrycky, Hébrard, Hinse, Mazeh, Orosz, Tsvetanov, \& Welsh}]{Kostov2014}
Kostov, V.~B., McCullough, P.~R., Carter, J.~A., {et~al.} 2014, \apj, 784, 14.
\newblock \url{http://stacks.iop.org/0004-637X/784/i=1/a=14}

\bibitem[{Kostov {et~al.}(2016)Kostov, Orosz, Welsh, Doyle, Fabrycky,
  Haghighipour, Quarles, Short, Cochran, Endl, Ford, Gregorio, Hinse, Isaacson,
  Jenkins, Jensen, Kane, Kull, Latham, Lissauer, Marcy, Mazeh, Müller, Pepper,
  Quinn, Ragozzine, Shporer, Steffen, Torres, Windmiller, \&
  Borucki}]{Kostov2016}
Kostov, V.~B., Orosz, J.~A., Welsh, W.~F., {et~al.} 2016, \apj, 827, 86.
\newblock \url{http://stacks.iop.org/0004-637X/827/i=1/a=86}

\bibitem[{{Kratter} \& {Shannon}(2014)}]{Kratter2014}
{Kratter}, K.~M., \& {Shannon}, A. 2014, \mnras, 437, 3727

\bibitem[{{Li} {et~al.}(2016){Li}, {Holman}, \& {Tao}}]{Li2016}
{Li}, G., {Holman}, M.~J., \& {Tao}, M. 2016, \apj, 831, 96

\bibitem[{{Martin}(2017)}]{Martin2017}
{Martin}, D.~V. 2017, \mnras, 465, 3235

\bibitem[{{Martin} {et~al.}(2015){Martin}, {Mazeh}, \& {Fabrycky}}]{Martin15}
{Martin}, D.~V., {Mazeh}, T., \& {Fabrycky}, D.~C. 2015, \mnras, 453, 3554

\bibitem[{{Martin} \& {Triaud}(2014)}]{Martin2014}
{Martin}, D.~V., \& {Triaud}, A.~H.~M.~J. 2014, \aap, 570, A91

\bibitem[{{Martin} \& {Triaud}(2015)}]{Martin2015}
---. 2015, \mnras, 449, 781

\bibitem[{{Martin} {et~al.}(2019){Martin}, {Triaud}, {Udry}, {Marmier},
  {Maxted}, {Collier Cameron}, {Hellier}, {Pepe}, {Pollacco}, {Segransan}, \&
  {West}}]{Martin2019}
{Martin}, D.~V., {Triaud}, A.~H.~M.~J., {Udry}, S., {et~al.} 2019, arXiv
  e-prints, arXiv:1901.01627

\bibitem[{{Mu{\~n}oz} \& {Lai}(2015)}]{Munoz15}
{Mu{\~n}oz}, D.~J., \& {Lai}, D. 2015, Proceedings of the National Academy of
  Science, 112, 9264

\bibitem[{{Musielak} {et~al.}(2005){Musielak}, {Cuntz}, {Marshall}, \&
  {Stuit}}]{Musielak2005}
{Musielak}, Z.~E., {Cuntz}, M., {Marshall}, E.~A., \& {Stuit}, T.~D. 2005,
  \aap, 434, 355

\bibitem[{{Orosz} \& {Hauschildt}(2000)}]{Orosz2000}
{Orosz}, J.~A., \& {Hauschildt}, P.~H. 2000, \aap, 364, 265

\bibitem[{{Orosz} {et~al.}(2019){Orosz}, {Welsh}, {Haghighipour}, {et al.}, {et
  al.}, \& {et al.}}]{Orosz2019}
{Orosz}, J.~A., {Welsh}, W.~F., {Haghighipour}, N., {et~al.} 2019, \aj, 157, 68

\bibitem[{Orosz {et~al.}(2012)Orosz, Welsh, Carter, Brugamyer, Buchhave,
  Cochran, Endl, Ford, MacQueen, Short, Torres, Windmiller, Agol, Barclay,
  Caldwell, Clarke, Doyle, Fabrycky, Geary, Haghighipour, Holman, Ibrahim,
  Jenkins, Kinemuchi, Li, Lissauer, Prša, Ragozzine, Shporer, Still, \&
  Wade}]{Orosz2012a}
Orosz, J.~A., Welsh, W.~F., Carter, J.~A., {et~al.} 2012, \apj, 758, 87.
\newblock \url{http://stacks.iop.org/0004-637X/758/i=2/a=87}

\bibitem[{{Orosz} {et~al.}(2012){Orosz}, {Welsh}, {Carter}, {Fabrycky},
  {Cochran}, {Endl}, {Ford}, {Haghighipour}, {MacQueen}, {Mazeh},
  {Sanchis-Ojeda}, {Short}, {Torres}, {Agol}, {Buchhave}, {Doyle}, {Isaacson},
  {Lissauer}, {Marcy}, {Shporer}, {Windmiller}, {Barclay}, {Boss}, {Clarke},
  {Fortney}, {Geary}, {Holman}, {Huber}, {Jenkins}, {Kinemuchi}, {Kruse},
  {Ragozzine}, {Sasselov}, {Still}, {Tenenbaum}, {Uddin}, {Winn}, {Koch}, \&
  {Borucki}}]{Orosz2012b}
{Orosz}, J.~A., {Welsh}, W.~F., {Carter}, J.~A., {et~al.} 2012, Science, 337,
  1511

\bibitem[{{Paardekooper} {et~al.}(2012){Paardekooper}, {Leinhardt},
  {Th{\'e}bault}, \& {Baruteau}}]{Paardekooper2012}
{Paardekooper}, S.-J., {Leinhardt}, Z.~M., {Th{\'e}bault}, P., \& {Baruteau},
  C. 2012, \apjl, 754, L16

\bibitem[{{P{\'a}l}(2012)}]{Pal2012}
{P{\'a}l}, A. 2012, \mnras, 420, 1630

\bibitem[{{Pilat-Lohinger} {et~al.}(2003){Pilat-Lohinger}, {Funk}, \&
  {Dvorak}}]{Pilat-Lohinger2003}
{Pilat-Lohinger}, E., {Funk}, B., \& {Dvorak}, R. 2003, \aap, 400, 1085

\bibitem[{{Quarles} \& {Lissauer}(2016)}]{Quarles2016}
{Quarles}, B., \& {Lissauer}, J.~J. 2016, \aj, 151, 111

\bibitem[{{Quarles} {et~al.}(2018){Quarles}, {Satyal}, {Kostov}, {Kaib}, \&
  {Haghighipour}}]{Quarles2018}
{Quarles}, B., {Satyal}, S., {Kostov}, V., {Kaib}, N., \& {Haghighipour}, N.
  2018, \apj, 856, 150

\bibitem[{{Rauer} {et~al.}(2014){Rauer}, {Catala}, {Aerts}, {Appourchaux},
  {Benz}, {Brandeker}, {Christensen-Dalsgaard}, {Deleuil}, {Gizon}, {Goupil},
  {G{\"u}del}, {Janot-Pacheco}, {Mas-Hesse}, {Pagano}, {Piotto}, {Pollacco},
  {Santos}, {Smith}, {Su{\'a}rez}, {Szab{\'o}}, {Udry}, {Adibekyan}, {Alibert},
  {Almenara}, {Amaro-Seoane}, {Eiff}, {Asplund}, {Antonello}, {Barnes},
  {Baudin}, {Belkacem}, {Bergemann}, {Bihain}, {Birch}, {Bonfils}, {Boisse},
  {Bonomo}, {Borsa}, {Brand{\~a}o}, {Brocato}, {Brun}, {Burleigh}, {Burston},
  {Cabrera}, {Cassisi}, {Chaplin}, {Charpinet}, {Chiappini}, {Church},
  {Csizmadia}, {Cunha}, {Damasso}, {Davies}, {Deeg}, {D{\'{\i}}az}, {Dreizler},
  {Dreyer}, {Eggenberger}, {Ehrenreich}, {Eigm{\"u}ller}, {Erikson}, {Farmer},
  {Feltzing}, {de Oliveira Fialho}, {Figueira}, {Forveille}, {Fridlund},
  {Garc{\'{\i}}a}, {Giommi}, {Giuffrida}, {Godolt}, {Gomes da Silva},
  {Granzer}, {Grenfell}, {Grotsch-Noels}, {G{\"u}nther}, {Haswell}, {Hatzes},
  {H{\'e}brard}, {Hekker}, {Helled}, {Heng}, {Jenkins}, {Johansen},
  {Khodachenko}, {Kislyakova}, {Kley}, {Kolb}, {Krivova}, {Kupka}, {Lammer},
  {Lanza}, {Lebreton}, {Magrin}, {Marcos-Arenal}, {Marrese}, {Marques},
  {Martins}, {Mathis}, {Mathur}, {Messina}, {Miglio}, {Montalban}, {Montalto},
  {Monteiro}, {Moradi}, {Moravveji}, {Mordasini}, {Morel}, {Mortier},
  {Nascimbeni}, {Nelson}, {Nielsen}, {Noack}, {Norton}, {Ofir}, {Oshagh},
  {Ouazzani}, {P{\'a}pics}, {Parro}, {Petit}, {Plez}, {Poretti}, {Quirrenbach},
  {Ragazzoni}, {Raimondo}, {Rainer}, {Reese}, {Redmer}, {Reffert},
  {Rojas-Ayala}, {Roxburgh}, {Salmon}, {Santerne}, {Schneider}, {Schou},
  {Schuh}, {Schunker}, {Silva-Valio}, {Silvotti}, {Skillen}, {Snellen}, {Sohl},
  {Sousa}, {Sozzetti}, {Stello}, {Strassmeier}, {{\v S}vanda}, {Szab{\'o}},
  {Tkachenko}, {Valencia}, {Van Grootel}, {Vauclair}, {Ventura}, {Wagner},
  {Walton}, {Weingrill}, {Werner}, {Wheatley}, \& {Zwintz}}]{Rauer14}
{Rauer}, H., {Catala}, C., {Aerts}, C., {et~al.} 2014, Experimental Astronomy,
  38, 249

\bibitem[{Rein \& Spiegel(2015)}]{Rein2015}
Rein, H., \& Spiegel, D.~S. 2015, \mnras, 446, 1424.
\newblock \url{http://dx.doi.org/10.1093/mnras/stu2164}

\bibitem[{{Sanz-Forcada} {et~al.}(2014){Sanz-Forcada}, {Desidera}, \&
  {Micela}}]{Sanz-Forcada2014}
{Sanz-Forcada}, J., {Desidera}, S., \& {Micela}, G. 2014, \aap, 570, A50

\bibitem[{Schwamb {et~al.}(2013)Schwamb, Orosz, Carter, Welsh, Fischer, Torres,
  Howard, Crepp, Keel, Lintott, Kaib, Terrell, Gagliano, Jek, Parrish, Smith,
  Lynn, Simpson, Giguere, \& Schawinski}]{Schwamb2013}
Schwamb, M.~E., Orosz, J.~A., Carter, J.~A., {et~al.} 2013, \apj, 768, 127.
\newblock \url{http://stacks.iop.org/0004-637X/768/i=2/a=127}

\bibitem[{{Silsbee} \& {Rafikov}(2015)}]{Silsbee2015}
{Silsbee}, K., \& {Rafikov}, R.~R. 2015, \apj, 808, 58

\bibitem[{{ter Braak}(2006)}]{Braak2006}
{ter Braak}, C. J.~F. 2006, Statistics and Computing, 16, 239.
\newblock \url{https://doi.org/10.1007/s11222-006-8769-1}

\bibitem[{Welsh {et~al.}(2012)Welsh, Orosz, Carter, Fabrycky, Ford, Lissauer,
  Prsa, Quinn, Ragozzine, Short, Torres, Winn, Doyle, Barclay, Batalha,
  Bloemen, Brugamyer, Buchhave, Caldwell, Caldwell, Christiansen, Ciardi,
  Cochran, Endl, Fortney, Gautier~III, Gilliland, Haas, Hall, Holman, Howard,
  Howell, Isaacson, Jenkins, Klaus, Latham, Li, Marcy, Mazeh, Quintana,
  Robertson, Shporer, Steffen, Windmiller, Koch, \& Borucki}]{Welsh2012}
Welsh, W.~F., Orosz, J.~A., Carter, J.~A., {et~al.} 2012, Nature, 481, 475 EP .
\newblock \url{http://dx.doi.org/10.1038/nature10768}

\bibitem[{{Welsh} {et~al.}(2015){Welsh}, {Orosz}, {Short}, {Cochran}, {Endl},
  {Brugamyer}, {Haghighipour}, {Buchhave}, {Doyle}, {Fabrycky}, {Hinse},
  {Kane}, {Kostov}, {Mazeh}, {Mills}, {M{\"u}ller}, {Quarles}, {Quinn},
  {Ragozzine}, {Shporer}, {Steffen}, {Tal-Or}, {Torres}, {Windmiller}, \&
  {Borucki}}]{Welsh2015}
{Welsh}, W.~F., {Orosz}, J.~A., {Short}, D.~R., {et~al.} 2015, \apj, 809, 26

\bibitem[{{Xu} \& {Lai}(2016)}]{Xu2016}
{Xu}, W., \& {Lai}, D. 2016, \mnras, 459, 2925

\end{thebibliography}



\end{document}